\documentclass[amssymb,aps,preprint,prb]{revtex4}
\usepackage{graphicx}

\begin{document}

\title{Effect of magnetic order on the superfluid response of single-crystal ErNi$_{2}$B$_{2}$C:
A penetration depth study}
\author{Elbert E. M. Chia}
\altaffiliation{Los Alamos National Laboratory, Los Alamos, New
Mexico 87545, USA}
\author{W. Cheong}
\author{Tuson Park}
\altaffiliation{Los Alamos National Laboratory, Los Alamos, New
Mexico 87545, USA}
\author{M. B. Salamon}
\affiliation{Department of Physics, University of Illinois at
Urbana-Champaign, 1110 W. Green Street, Urbana Illinois 61801}
\author{Eun-Mi Choi}
\author{Sung-Ik Lee}
\altaffiliation{Quantum Materials Research Laboratory, Korea Basic
Science Institute, Daejeon 305-333, Republic of Korea}
\affiliation{National Creative Research Initiative Center for
Superconductivity and Department of Physics, Pohang University of
Science and Technology, Pohang 790-784, Republic of Korea}
\date{\today}

\begin{abstract}
We report measurements of the in-plane magnetic penetration depth
$\Delta \lambda $(T) in single crystals of ErNi$_{2}$B$_{2}$C down
to $\sim$0.1~K using a tunnel-diode based, self-inductive technique
at 21~MHz. We observe four features: (1) a slight dip in $\Delta
\lambda $(T) at the N$\acute{e}$el temperature $T_{N}$ = 6.0~K, (2)
a peak at $T_{WFM}$ = 2.3~K, where a weak ferromagnetic component
sets in, (3) another maximum at 0.45~K, and (4) a final broad drop
down to 0.1~K. Converting to superfluid density $\rho _{s}$, we see
that the antiferromagnetic order at 6~K only slightly depresses
superconductivity. We seek to explain some of the above features in
the context of antiferromagnetic superconductors, where competition
between the antiferromagnetic molecular field and spin fluctuation
scattering determines increased or decreased pairbreaking.
Superfluid density data show only a slight decrease in pair density
in the vicinity of the 2.3~K feature, thus supporting other
evidences against bulk ferromagnetism in this temperature range.

\end{abstract}

\maketitle The magnetic members of the rare-earth (RE) nickel
borocarbide family, RENi$_{2}$B$_{2}$C (RE = Ho, Er, Dy, etc) have
attracted much interest due to the interplay between magnetism and
superconductivity. ErNi$_{2}$B$_{2}$C, in particular, is a good
candidate for study: superconductivity starts at $T_{c}$
$\approx$~11~K, before antiferromagnetic (AF) order sets in
\cite{Cho95} at $T_{N}$ $\approx$~6~K. In the AF state the Er spins
are directed along the {\it b}-axis, forming a transversely
polarized, incommensurate spin-density-wave (SDW) state, with
modulation vector modulation vector \cite{Zarestky95} $\delta $ =
0.553$a^{\ast}$ (a$^{\ast}$ = 2$\pi $/{\it a}), before squaring up
at lower temperatures \cite{Choi01}. Below $T_{WFM}$=2.3~K a net
magnetization appears, superposed on a modulation with a periodicity
of 20$a^{\ast}$, confirming the microscopic existence of spontaneous
weak ferromagnetism (WFM) with superconductivity
\cite{Choi01,Canfield96}.

The fact that $T_{N} < T_{c}$ enables us to study the influence of
magnetism on superconductivity. In particular, in this paper we
study, via the penetration depth, the pairbreaking effects of the
various magnetic orders on the superfluid response of this material.
There have been several previous penetration depth measurements on
ErNi$_{2}$B$_{2}$C. Jacobs {\it et al.} \cite{Jacobs95} measured the
microwave surface impedance of single-crystal ErNi$_{2}$B$_{2}$C
from $T_{c}$ down to 4~K, but did not see the AF transition at 6~K.
They concluded that the AF transition is not accompanied by changes
in pairbreaking in zero field. Andreone {\it et al.} measured the
microwave properties of ErNi$_{2}$B$_{2}$C thin films
--- microwave surface resistance \cite{Andreone99a} down to 2~K,
and the change in penetration depth \cite{Andreone98} from 2--5~K.
They too, saw no feature at {\it T}$_{N}$, and attributed its
absence to the smearing of the susceptibility $\chi $({\it T}).
Gammel {\it et al.} performed small-angle neutron measurements
(SANS) \cite{Gammel99} on single-crystal ErNi$_{2}$B$_{2}$C down to
4~K, finding a decrease of $\lambda $ below $T_{N}$ that they could
not account for quantitatively. In this paper we present
high-precision measurements of the in-plane magnetic penetration
depth of single-crystal ErNi$_{2}$B$_{2}$C down to 0.1~K. We see
features at $T_{N}$ and $T_{WFM}$, and ascribe these to the
pairbreaking effects of AF order at $T_{N}$ = 6~K and the weak
ferromagnetic ordering at $T_{WFM}$ = 2.3~K. We also observe a peak
at 0.45~K, which we attribute to the presence of a spontaneous
vortex phase \cite{Chia04b} (SVP), expected to occur in
superconductors where ferromagnetism and superconductivity coexist
\cite{Blount79,Tachiki79,Greenside81}. The superfluid density graph
indicates that these three magnetic orderings coexist with
superconductivity, i.e. they do not destroy superconductivity in
this material.

Various theories of antiferromagnetic superconductors have been
proposed
\cite{Baltensperger86,Zwicknagl81,Machida80b,Ramakrishnan81,Nass81,Nass82,Chi92}.
We shall follow that of Chi and Nagi \cite{Chi92}, which is an
extension of the mean-field model by Nass {\it et al.}
\cite{Nass81,Nass82} to the regime where the superconducting gap
$\Delta $ is finite, and it includes the effects of
spin-fluctuations, molecular field and impurities. In the Chi-Nagi
(CN) model, which applies specifically to superconductors with
$T_{N} < T_{c}$, two temperature regimes are separately considered.
First, in the paramagnetic regime ($T_{N} < T < T_{c}$), the
depression of $T_{c}$ with respect to the non-magnetic counterparts,
LuNi$_{2}$B$_{2}$C or YNi$_{2}$B$_{2}$C, is due to the exchange
scattering of the conduction electrons from the spins of the RE Er
ions. Assuming that the exchange interaction is weak, this
paramagnetic phase of ErNi$_{2}$B$_{2}$C can be accounted for by the
Abrikosov-Gorkov (AG) pairbreaking theory \cite{Abrikosov61}.
Second, in the AF phase ($T < T_{N}$), the effect of pairbreaking
depends on the competition \cite{Ro84} between the
temperature-dependent AF molecular field (with parameter $H_{Q}(T)$)
and spin-fluctuation scattering of the conduction electrons, the
latter by both magnetic RE ions (parameter 1/$\tau_{2}^{eff}$) and
non-magnetic impurities (parameter 1/$\tau_{1}$). The molecular
field opens AF gaps on parts of the Fermi surface (FS), hence
destroying the superconducting gap in those areas. The non-magnetic
impurities do not affect the BCS state for an $s$-wave
superconductor \cite{Anderson59}, but weaken the effect of the AF
field by destroying the pairing state for charge density waves or
spin density waves \cite{Zittartz67} --- thus \textit{non-magnetic
impurities promote the recovery of superconductivity}. Moreover, the
effect of the molecular field and spin fluctuations is governed by a
sum rule \cite{Ro84}, and the competition between them determines
whether the AF phase gives increased or decreased pairbreaking below
$T_{N}$. The total electronic effective magnetic scattering rate
1/$\tau_{2}^{eff}$ is temperature-dependent and decreases with
decreasing temperature (as the magnetic moments become more and more
frozen). The assumptions of the CN model are: (1) the effect of
inelastic scattering, which is relevant only for $T \ll T_{N}$, can
be ignored; (2) BCS $s$-wave pairing and (3) a one-dimensional (1-D)
electron band that satisfies the nesting condition $\epsilon_{k} =
-\epsilon_{k+Q}$.

The following equations of the CN model were used \cite{Chi92}.
The temperature-dependence of the superconducting gap is
determined from
\begin{widetext}
\begin{equation}
\mbox{(AG equation)} \qquad \ln
\left(\frac{T_{c}}{T_{c0}}\right)=\psi
\left(\frac{1}{2}\right)-\psi \left(\frac{1}{2}+\frac{1}{2\pi
T_{c}\tau _{2}^{eff}}\right),
\label{eqn:AG}
\end{equation}
\end{widetext}

\begin{widetext}
\begin{equation}
\mbox{(Renormalized frequency)} \qquad \widetilde{\omega }_{n\pm
}=\omega _{n}+Y_{\mp }\frac{\widetilde{\omega }_{n+}}{2\lambda
_{+}}+ Y_{\pm }\frac{\widetilde{\omega }_{n-}}{2\lambda _{-}},
\label{eqn:omegan}
\end{equation}
\end{widetext}

\begin{widetext}
\begin{equation}
\mbox{(Renormalized gap)} \qquad \widetilde{\Delta }_{n\pm
}=\Delta \pm H_{Q}(T)+X_{\mp }\frac{\widetilde{\Delta
}_{n+}}{2\lambda _{+}} +X_{\pm }\frac{\widetilde{\Delta
}_{n-}}{2\lambda _{-}},
\label{eqn:Deltan}
\end{equation}
\end{widetext}

\begin{widetext}
\begin{equation}
\mbox{(Gap equation)} \qquad \ln \frac{T}{T_{c0}}=\pi T\sum
\left\{\frac{1}{\Delta
}\left[\frac{1}{(U_{n+}^{2}+1)^{1/2}}+\frac{sgn
(U_{n-})}{(U_{n-}^{2}+1)^{1/2}}\right]-\frac{2}{\omega
_{n}}\right\}, \label{eqn:gapeqn}
\end{equation}
\end{widetext} where $T_{c0}$ is the superconducting transition
temperature of the non-magnetic member of the borocarbide family
LuNi$_{2}$B$_{2}$C or YNi$_{2}$B$_{2}$C, $\psi$ is the digamma
function, X$_{\pm}$ and Y$_{\pm}$ are linear combinations of the
magnetic (1/$\tau_{2}^{eff}$), non-magnetic (1/$\tau_{1}$) and
spin-orbit (1/$\tau_{so}$), scattering rates
\begin{widetext}
\begin{equation}
\qquad Y_{\pm} = \frac{1}{2}\left(\frac{1}{\tau_{1}}+
\frac{1}{\tau_{2}^{eff}} + \frac{1}{\tau_{so}}\right) \pm
\frac{1}{2}\left(\frac{1}{\tau_{1}}+ \frac{1}{3\tau_{2}^{eff}} +
\frac{1}{3\tau_{so}}\right),
\label{eqn:Ypm}
\end{equation}
\end{widetext}

\begin{widetext}
\begin{equation}
\qquad X_{\pm} = \frac{1}{2}\left(\frac{1}{\tau_{1}} -
\frac{1}{\tau_{2}^{eff}} + \frac{1}{\tau_{so}}\right) \pm
\frac{1}{2}\left(\frac{1}{\tau_{1}}+ \frac{1}{3\tau_{2}^{eff}} -
\frac{1}{3\tau_{so}}\right).
\label{eqn:Xpm}
\end{equation}
\end{widetext} $\lambda_{\pm}$ = [$\widetilde{\omega}_{n
\pm}^{2}$ + $\widetilde{\Delta}_{n \pm}^{2}$]$^{1/2}$; {\it U}$_{n
\pm}$ = $\widetilde{\omega}_{n \pm}$/$\widetilde{\Delta}_{n \pm}$;
$\omega_{n}$ = $\pi{\it T}$(2n+1) is the Matsubara frequency. In the
paramagnetic phase the distinction between $+$ and $-$ is lost,
giving $\widetilde{\omega}_{n \pm} \equiv \widetilde{\omega}_{n}$,
$\widetilde{\Delta}_{n \pm} \equiv \widetilde{\Delta}_{n}$ and so
{\it U}$_{n}$ = $\widetilde{\omega}_{n}$/$\widetilde{\Delta}_{n}$.

The temperature-dependence of the superfluid density $\rho_{s}$ is
given by \cite{Chi92}
\begin{equation}
\qquad \rho _{s}(T)\equiv \left[ \frac{\lambda ^{2}(0)}{\lambda
^{2}(T)} \right] ^{2}=\left[ \pi T\sum\limits_{n\geq 0}A(\omega
_{n})\right],
\label{eqn:AFrho}
\end{equation} where
\begin{widetext}
\begin{equation} \label{eqn:Aomegan}
\quad A(\omega _{n})=\frac{\widetilde{\Delta
}_{n+}^{2}-\widetilde{\omega }_{n+}^{2}}{4\varepsilon _{1}^{3}}
+\frac{\widetilde{\Delta }_{n-}^{2}-\widetilde{\omega
}_{n-}^{2}}{4\varepsilon _{2}^{3}}+\frac{1} {4\varepsilon
_{1}}+\frac{1}{4\varepsilon _{2}}+\frac{1}{\varepsilon
_{1}+\varepsilon _{2}} + \frac{\widetilde{\Delta
}_{n+}\widetilde{\Delta }_{n-}-\widetilde{\omega
}_{n+}\widetilde{\omega }_{n-}} {\varepsilon _{1}\varepsilon
_{2}(\varepsilon _{1}+\varepsilon _{2})},
\end{equation}
\end{widetext}
\begin{equation} \label{eqn:epsilon12}
\varepsilon_{1} =\mid (\widetilde{\omega }_{n+}^{2}+
\widetilde{\Delta }_{n+}^{2})^{1/2}\mid, \varepsilon_{2} =\mid
(\widetilde{\omega }_{n-}^{2}+ \widetilde{\Delta
}_{n-}^{2})^{1/2}\mid.
\end{equation}

In the paramagnetic phase ({\it T}$_{N}$ $<$ {\it T} $<$ {\it
T}$_{c}$), $\rho_{s}$ is given by (P: paramagnetic)
\begin{equation}
\qquad \rho _{s}^{P}(T)=\left[ 2\pi T\sum\limits_{n\geq 0}\frac{1}
{\varepsilon (1+U_{n}^{2})}\right],
\label{eqn:AGrho}
\end{equation} where $\varepsilon =\mid (\widetilde{\omega
}_{n}^{2}+ \widetilde{\Delta }_{n}^{2})^{1/2}\mid$. Note that this
expression for the superfluid density is for materials in the dirty
limit. This is consistent with the Er ions in ErNi$_{2}$B$_{2}$C
being the "impurity" ion when compared to either LuNi$_{2}$B$_{2}$C
or YNi$_{2}$B$_{2}$C.

We turn next to the parameters of the model. The effective
magnetic scattering rate (1/$\tau_{2}^{eff}$) from RE ions
(1/$\tau_{2}^{R}$) and magnetic impurities (1/$\tau_{2}^{i}$) is
given by
\begin{equation}
\qquad \frac{1}{\tau_{2}^{eff}}=\left\{
\begin{array}{ll}
\frac{1}{\tau_{2}^{i}}+\frac{1}{\tau_{2}^{R}}  & (T > T_{N}) \\
\frac{1}{\tau_{2}^{i}}+\frac{1}{\tau_{2}^{R}}(1-F^{2}(T)) & (T
    \leq T_{N})
\end{array}
\right., \label{eqn:mageffscatterT}
\end{equation}

\begin{equation}
\mbox{where} \qquad \frac{1}{\tau _{2}^{R}}=2\pi
n_{R}N(0)J(J+1)(g_{J}-1)^{2}I^{2}.
\label{eqn:magscatter0}
\end{equation}

The AF molecular field is given by
\begin{equation}
\qquad H_{Q}(T)=H_{Q}(0)F(T),
\label{eqn:HQT}
\end{equation}
\begin{equation}
\mbox{where} \qquad H_{Q}(0)=n_{R}I\mid g_{J}-1\mid \sqrt{J(J+1)}.
\label{eqn:HQ0}
\end{equation} $n_{R}$ is the concentration of RE
ions, $I$ is the exchange energy, $g_{J}$ is the Land$\acute{e}$
factor, and $J$ is the total angular momentum of the RE ion. The
function $F(T)$ can be approximated by the empirical relation
\begin{equation}
\qquad F(T)=1-\left(\frac{T}{T_{N}}\right)^{\nu },
\label{eqn:FT}
\end{equation} where $\nu$ is a parameter obtained by fitting $F(T)$ to
sublattice magnetization data.

The values of the renormalized frequencies $\widetilde{\omega}_{n
\pm}$ and gaps $\widetilde{\Delta}_{n \pm}$ are determined
self-consistently: for a fixed temperature $T$ and Matsubara index
$n$, one determines $\widetilde {\omega}_{n \pm}$ and $\widetilde
{\Delta}_{n \pm}$ from Equations~\ref{eqn:omegan} and
\ref{eqn:Deltan} such that they also satisfy Eqn.~\ref{eqn:gapeqn}.
After computing the $\widetilde {\omega}_{n \pm}$'s and $\widetilde
{\Delta}_{n \pm}$'s for a fixed $T$, one then substitutes these
values into Eqn.~\ref{eqn:AFrho} or \ref{eqn:AGrho} to obtain the
superfluid density $\rho_{s}$ at that temperature $T$.

Details of sample growth and characterization are described in
Ref.~\onlinecite{Cho95}. The samples were then annealed according
to conditions described in Ref.~\onlinecite{Miao02}. The
superconducting transitions of our sample were measured by
low-field ($H$=5~G) magnetization, zero-field resistivity and
zero-field specific-heat measurements. From magnetization data,
the onset of superconducting diamagnetism appears at $T$=11.0~K
and 90$\%$ of the full diamagnetic magnetization is reached at
$T$=9.6~K. Resistivity data show a superconducting onset at a
higher temperature of 11.3~K and zero resistivity at 9.6~K. The
mid-point of the specific-heat jump \cite{Tuson02} yields a
$T_{c}$ of 10.1~K. A comparison of the three measurements show
that bulk superconductivity occurs at $T_{c} \approx$~10~K,
whereas the initial decrease of resistivity at $\sim$11~K may be
due to some sort of filamentary superconductivity.

The parameters of this model are determined as follows. We denote
$\Delta_{0}$ and $\Delta$(0) to be the zero-temperature
superconducting gap amplitude of YNi$_{2}$B$_{2}$C and
ErNi$_{2}$B$_{2}$C, respectively. Tunneling measurements
\cite{Rybaltchenko96} yield $\Delta_{0} = 1.83T_{c0}$. From the
experimental values of $T_{c}$ (10.1~K for ErNi$_{2}$B$_{2}$C) and
$T_{c0}$ (15.5~K for YNi$_{2}$B$_{2}$C), Eqn.~\ref{eqn:AG} gives
1/$\tau_{2}^{eff} \Delta_{0}$ = 0.227. We assume 1/$\tau_{2}^{i}$ =
0, in which case Eqn.~\ref{eqn:mageffscatterT} gives 1/$\tau_{2}^{R}
\Delta_{0}$=0.227. For ErNi$_{2}$B$_{2}$C, using the values
$n_{R}$=1/6, $J$=7.5, $g_{J}$=1.2, $N(0)$=0.36 states/eV-atom-spin
\cite{Dugdale99}, we obtain $I$=0.024~eV from
Eqn.~\ref{eqn:magscatter0} which is comparable with the experimental
value of 0.031 eV \cite{Hagary00}. This justifies our assuming
1/$\tau_{2}^{i}$ = 0, as any finite $\tau_{2}^{i}$ would make $I$
even smaller than the experimental value. Equation~\ref{eqn:HQ0}
then gives H$_{Q}(0)/\Delta_{0}$=2.6. From the
temperature-dependence of the magnetic Bragg peak intensity
\cite{Lynn97,Choi01} below $T_{N}$ we obtain $\nu $ = 4.8 in
Eqn.~\ref{eqn:FT}. The only remaining free parameter of the theory
is 1/$\tau$, the non-magnetic scattering rate, defined to be $1/\tau
= 1/\tau_{1} + 2/3\tau_{so}$. In an AFSC, usually $1/\tau \gg
1/\tau_{so}$ \cite{Chi92}, so for the present study we assume
$1/\tau_{so}$=0, such that $1/\tau = 1/\tau_{1}$.

To see the pairbreaking effects of the various magnetic orders we
need to convert $\Delta \lambda (T)$ to $\rho_{s} (T)$, the
superfluid density. To determine $\rho_{s} (T)$ we need the value
of $\lambda $(0), which has been reported over a range
\cite{Gammel99,Cho95} from ~700 \AA\ to 1150 \AA. We take $\lambda
$(0) to be a parameter in our model, keeping in mind that it has
to be in the vicinity of the above two values.

\begin{figure}
\centering
\includegraphics[width=16cm,clip]{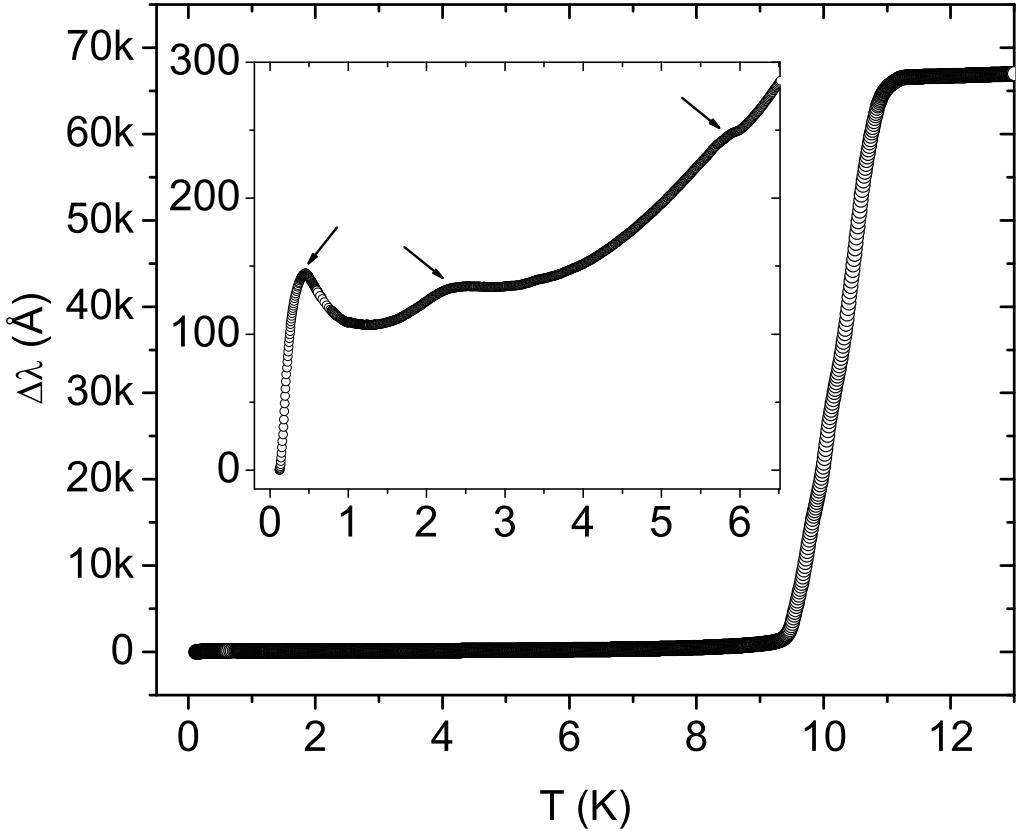}
\caption{($\bigcirc$) Temperature dependence of the penetration
depth $\Delta \lambda (T)$ from $\sim$0.1~K to 13K. Inset: $\Delta
\lambda (T)$ below 6.5~K. The arrows show features at 6~K (AFM),
2.3~K (WFM) and 0.45~K (SVP).}\label{fig:Lambda}
\end{figure}

Figure~\ref{fig:Lambda} shows the temperature-dependence of the
in-plane penetration depth $\Delta \lambda (T)$. The onset of
superconductivity, $T_{c}^{\ast}$, is 11.3~K, showing that this is a
high-quality single crystal. We also see the following features: (1)
a slight dip in $\Delta \lambda (T)$ at $T_{N}$=6.0~K, (2) a peak at
$T_{WFM}$=2.3~K, (3) another maximum at 0.45~K, and (4) a final
broad drop down to 0.1~K. We attribute the last two features to the
presence of the SVP. For dirty AF superconductors, the penetration
depth is expected to decrease below $T_{N}$ by both \cite{Gray83}
the susceptibility ($\chi $) and mean free path ({\it l}) as
$\lambda \sim \lambda_{L}^{\prime}/\sqrt{1+4 \pi \chi}$, where
$\lambda_{L}^{\prime} \approx \lambda_{L}(1+\xi_{0}/l)$. Neither
effect, however, explains our data: First, using the mean-field
expression for $\chi $, in order to reproduce the experimental dip,
the peak in $\chi $ at $T_{c}$ has to be at least an order of
magnitude larger than that suggested by magnetization measurements
\cite{Cho95}. Second, from our resistivity data we obtained
$H_{c2}(T)$, and hence we calculated $\xi_{0}(T)$, $l(T)$, and
lastly, $\lambda_{L}^{\prime}$. Our values of $\lambda_{L}^{\prime}$
also are unable to explain the magnitude of the drop of $\lambda$
below $T_{N}$ --- a conclusion also shared by Gammel \textit{et al.}
from their SANS data \cite{Gammel99}.

\begin{figure}
\centering
\includegraphics[width=16cm,clip]{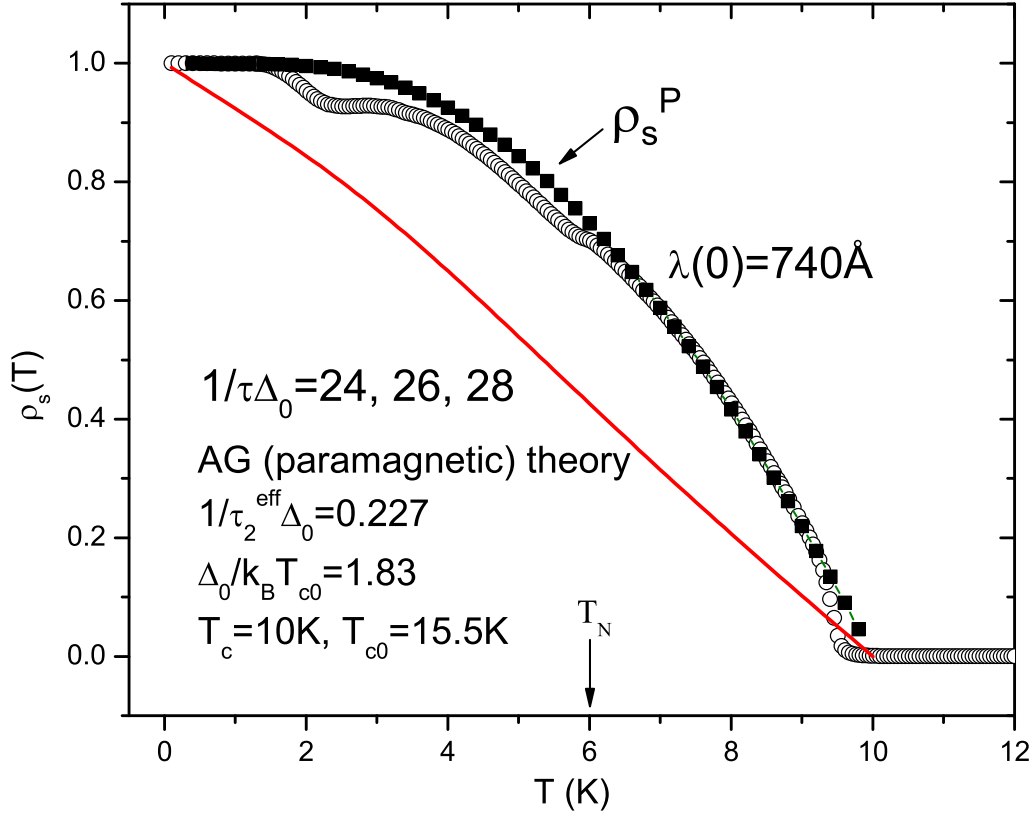}
\caption{($\bigcirc$) Experimental superfluid density $\rho_{s}(T)$
= [$\lambda^{2}$(0)/$\lambda^{2}(T)$] calculated from $\Delta
\lambda (T)$ data in Fig.~\ref{fig:Lambda}, from 0.1~K to $T_{c}$.
Solid squares = Calculated $\rho_{s}(T)$ assuming paramagnetic phase
from $T$=0 to $T_{c}$. Note that $1/\tau \Delta_{0}$=24, 26 or 28
give virtually the same theoretical curve. Solid line: Theoretical
curve for $d_{x^{2}-y^{2}}$ order parameter. The arrow denotes
N$\acute{e}$el temperature at 6~K.}\label{fig:ENBCrhoAG}
\end{figure}

Since the CN model does not take into account the effect of the SVP
on the superfluid density, we neglect it when we convert to
superfluid density $\rho_{s}$: First we assume $\Delta \lambda (T)$
follows a power-law temperature dependence at low temperatures from
the combination of gap-minima observed in non-magnetic borocarbides
and the increased pairbreaking as Er spins disorder. Consequently,
we set $\lambda_{low} (T) = \lambda (0) (1+bT^{2})$ with
$b=0.036$~K$^{-2}$ from Ref.~\onlinecite{Chia04b}. Next we offset
$\lambda_{low}$ until it matches the data at 1.3~K, the local
minimum in $\Delta \lambda$ in the vicinity of 1.5~K. Finally we
convert $\Delta \lambda$ to $\rho_{s}$ in Fig.~\ref{fig:ENBCrhoAG}.

The superfluid data leads to some important observations. First, the
data in the paramagnetic phase ($T > T_{N}$) fit the theoretical
curve based on an isotropic superconducting gap (solid squares), and
not that based on nodes. The solid line shows a superfluid
calculation based on a $d_{x^{2}-y^{2}}$ order parameter. Second,
the superconductivity is only slightly depressed in the AF phase
below $T_{N}$. The best fit to data above $T_{N}$ (solid squares) is
obtained when $\lambda$(0) = 740 \AA\ --- here we assume
paramagnetism from $T$= 0 to $T_{c}$, neglecting AF order, with
parameter 1/$\tau \Delta_{0}$ = 24. The paramagnetic curve is almost
unchanged if one uses 1/$\tau \Delta_{0}$=26 or 28. We see that the
paramagnetic curve fits the data above $T_{N}$, and overestimates
the data in the AF phase. We will show below that, as one crosses
$T_{N}$ from above, the AF phase leads to increased or decreased
superfluid density depending on the combined effects of the
following three factors: (1) the AF molecular field, which decreases
the magnitude of the superconducting gap and hence decreases the
superfluid density, (2) freezing out of spin fluctuations, leading
to decreased pairbreaking and hence increased superfluid density,
and (3) scattering from non-magnetic impurities, which reduces the
suppression of the gap by the molecular field. The parameters of
ErNi$_{2}$B$_{2}$C are such that these three effects result in a
slight decrease in superfluid density below $T_{N}$.

\begin{figure}
\centering
\includegraphics[width=16cm,clip]{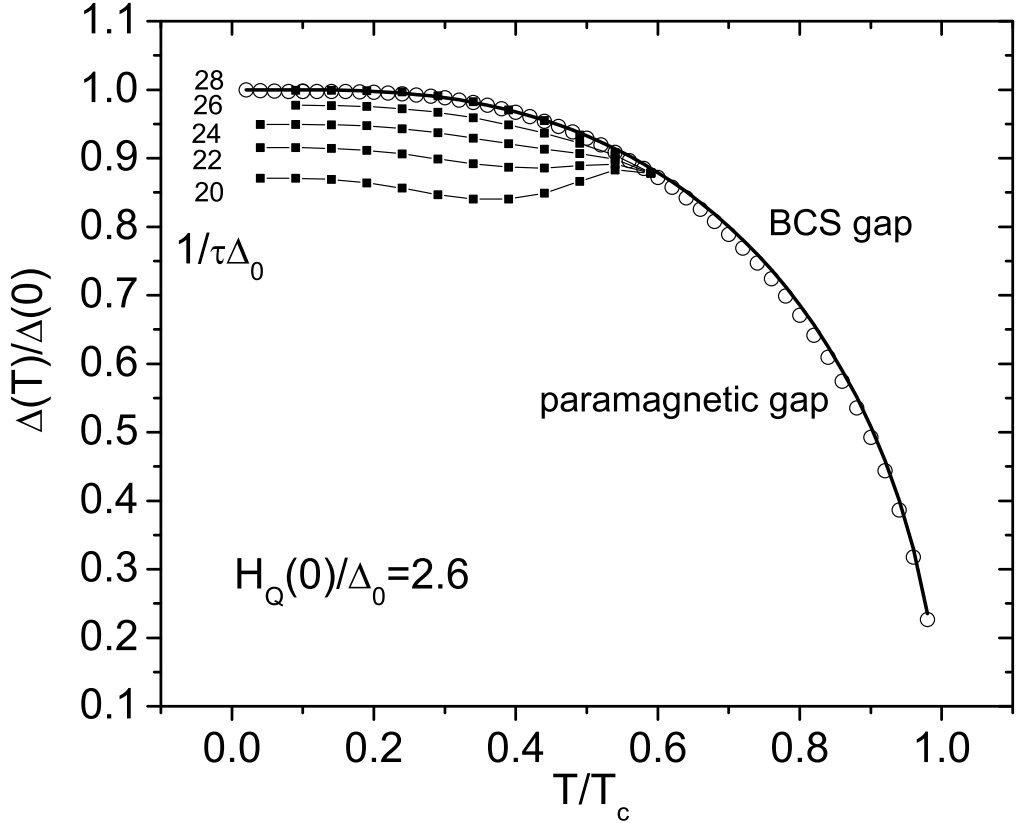}
\caption{Chi-Nagi model calculation for the superconducting gap
$\Delta$. Solid line = BCS temperature-dependence. ($\bigcirc$)
Paramagnetic gap. Solid squares = Incorporating AF phase below
$T_{N}$, for different values of 1/$\tau \Delta_{0}$. Note that when
1/$\tau \Delta_{0}$=28, superconductivity is fully recovered.
}\label{fig:ENBCrhoCN}
\end{figure}

\begin{figure}
\centering
\includegraphics[width=16cm,clip]{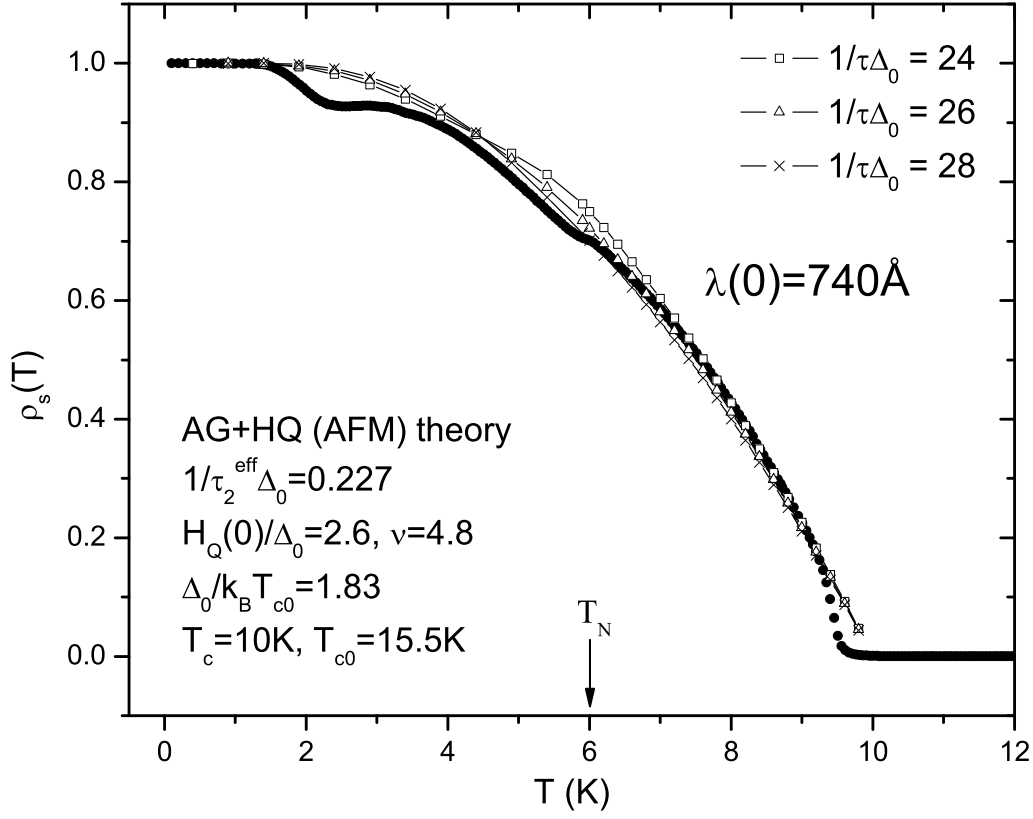}
\caption{Superfluid density $\rho_{s}(T)$=
[$\lambda^{2}$(0)/$\lambda^{2}(T)$] from 0.1~K to $T_{c}$. Solid
circles = data. AF-phase fitting curves for 1/$\tau \Delta_{0}$=24
($\Box$), 26 ($\bigtriangleup$), 28
($\times$).}\label{fig:ENBCrhoHQ}
\end{figure}

Fig.~\ref{fig:ENBCrhoCN} shows the calculated superconducting gap
amplitude $\Delta (T)$, in the presence of $H_{Q}(T)$, for various
values of 1/$\tau \Delta_{0}$, as well as the paramagnetic curve.
The normalized paramagnetic gap (open circles) agrees excellently
with the BCS gap (solid line). Tunneling measurements
\cite{Yanson00} also show that $\Delta(T)$ follows the BCS curve
above $T_{N}$. Next, as shown in Fig.~\ref{fig:ENBCrhoCN}, in the AF
phase, as 1/$\tau \Delta_{0}$ increases, superconductivity is
gradually recovered, as evidenced by the increase of $\Delta (T)$
(solid squares). Fig.~\ref{fig:ENBCrhoHQ} shows $\rho_{s} (T)$ for
various values of 1/$\tau \Delta_{0}$. Notice that when 1/$\tau
\Delta_{0}$=26, (1) $\rho_{s}$ decreases only slightly at $T_{N}$,
and (2) $\Delta$ is only slightly depressed below the BCS value, in
agreement with tunneling data \cite{Yanson00,Watanabe00b}. This
value of 1/$\tau \Delta_{0}$ corresponds to a mean free path (mfp)
of 45~\AA. Though this curve still overestimates the experimental
data below $T_{N}$, it at least fits the data better than the
paramagnetic curve.

To see if this value of 1/$\tau \Delta_{0}$ is reasonable, we take
$\Delta(0) = 1.83 T_{c}$ for ErNi$_{2}$B$_{2}$C
\cite{Rybaltchenko96}, for which the BCS coherence length
$\xi_{0BCS}^{\Delta}$ = $\hbar v_{F}$/$\pi \Delta$(0) = 470 \AA,
where {\it v}$_{F}$ = 3.6 $\times$ 10$^{5}$ m/s is taken from
band-structure calculations \cite{Pickett94} for
LuNi$_{2}$B$_{2}$C and YNi$_{2}$B$_{2}$C. Using the relation
\begin{equation} \label{eqn:Hc20}
H_{c2}(0) = 0.693 T_{c}(\frac{dH_{c2}}{dT}) \Big|_{T_{c}},
\end{equation}
and ${dH_{c2}/dT |_{T_{c}}}$ ($H \parallel c$) = $-$2.67~kOe/K
\cite{Cho95}, we obtain the coherence length
\begin{equation} \label{eqn:xi0Hc20}
\xi_{0}^{H_{c2}} = \sqrt{\frac{\phi_{0}}{2 \pi H_{c2}(0)}} = 130
\mbox{ \AA}.
\end{equation}
Finally, using the relation \cite{Rathnayaka97}
\begin{equation} \label{eqn:xi0BCS}
\xi_{0}^{H_{c2}} = 0.85(\xi_{0BCS}^{\Delta} l)^{1/2},
\end{equation}
we obtain the mfp $l$=42~\AA. On the other hand, from the
resistivity value just above $T_{c}$, $\rho(T_{c}^{\ast})$=5.8 $\mu
\Omega$-cm, we get $l$=56~\AA. These two values agree well with the
value of 45~\AA\ calculated from 1/$\tau \Delta_{0}$=26 obtained
earlier, implying that this particular value of the non-magnetic
scattering rate needed to explain our $\rho_{s}$ data is consistent
with other measurements. Note that this value of mfp calculated from
$H_{c2}$ data does not depend on the exact value of $T_{c}$. Also,
the prefactors 0.693 and 0.85 in the above relations are for
materials in the dirty limit. Here $l < \xi_{0}$, so our
ErNi$_{2}$B$_{2}$C sample may be considered as ``quasidirty". It is
puzzling that our sample has a high $T_{c}$ and be considered
quasi-dirty, yet this is consistent with the results of other
papers. Also, in this sample the non-magnetic scattering rate
(1/$\tau $) is at least two orders of magnitude larger than the
effective magnetic scattering rate (1/$\tau_{2}^{eff}$), thus the
mfp value is largely determined by 1/$\tau $. Our mfp value,
however, is smaller than the 90~\AA\ obtained from resistivity
measurements just above $T_{c}$ in Ref.~\onlinecite{Bhatnagar97}.
The CN model is thus able to explain our superfluid density data,
both qualitatively and quantitatively. Our data, in agreement with
others, also shows that AF order coexists with superconductivity
below $T_{N}$.

It is also interesting to note that according to
Ref.~\onlinecite{Ro84}, a near-exact cancellation of
spin-fluctuation and molecular-field effects occur at a critical
value of $N(0)J^{cf}$$\sim$1.0 $\times$ 10$^{-3}$, where
$J^{cf}$=$I\mid g_{J}-1\mid$ is the conduction-electron local
(\textit{f}) spin exchange. For the case of ErNi$_{2}$B$_{2}$C, we
obtain $N(0)J^{cf} = 1.7 \times 10^{-3}$, which explains the small
change in pairbreaking at $T_{N}$.

We turn next to an alternative explanation for the change of
$\rho_{s}$ at $T_{N}$. Ramakrishnan and Varma \cite{Ramakrishnan81}
predicted that for materials with a nested FS, since the peak in
susceptibility and the joint density of states (defined as the
difference between the susceptibility in the superconducting state
and the normal state) occur at the same $Q$-value, one should expect
an increase in pairbreaking at $T_{N}$. Conversely, a non-nested FS
will give rise to decreased pairbreaking at $T_{N}$. Two-dimensional
angular correlation of electron-positron annihilation radiation
measurements show that only one out of the three FS sheets in
LuNi$_{2}$B$_{2}$C possesses nesting properties, thereby accounting
for the propensity for magnetic ordering found in the other magnetic
members of the RE nickel borocarbides \cite{Dugdale99}. Also,
Dugdale {\it et al.} \cite{Dugdale99} estimated that the fraction of
the FS that would be able to participate in nesting is only 4.4$\%$.
Contrast this with CN model, which assumes perfect 1-D nesting.
Hence the increased pairbreaking due to partial nesting on one FS
sheet is partially compensated by decreased pairbreaking by the
other two sheets, resulting in only a slight increase in
pairbreaking at $T_{N}$.

\begin{figure}
\centering
\includegraphics[width=16cm,clip]{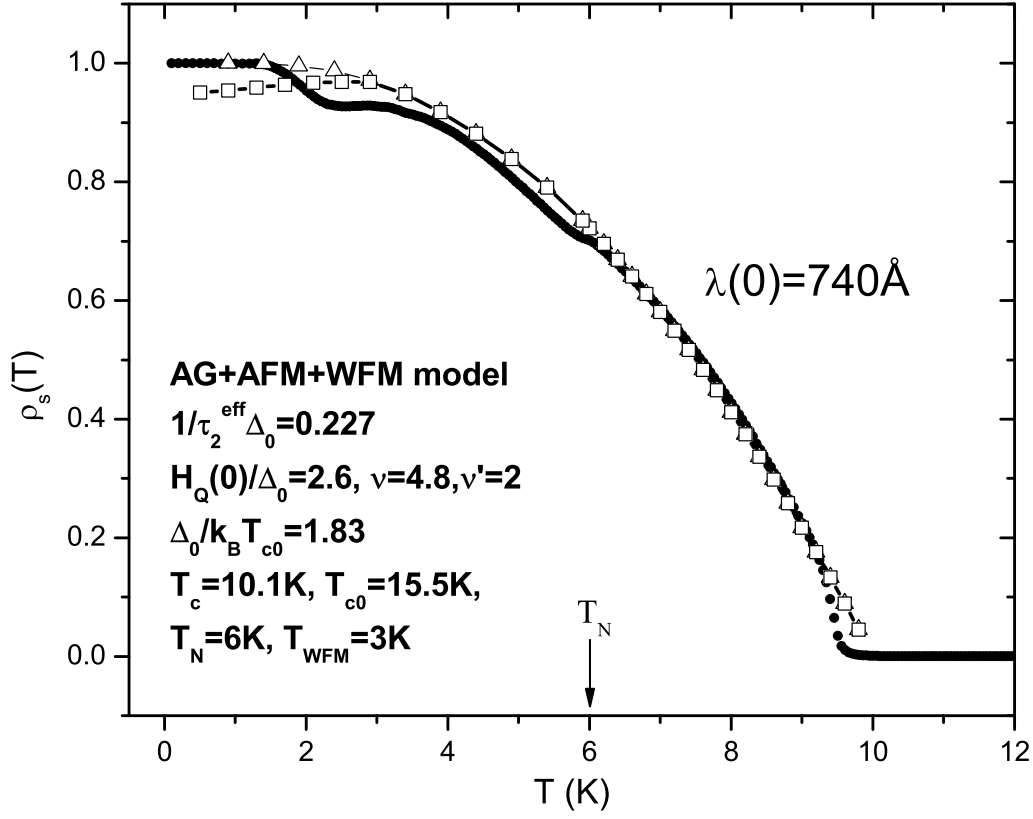}
\caption{$\rho_{s}(T)$ from 0.1~K to $T_{c}$. ($\bigcirc$) Data.
($\bigtriangleup$) Calculated AF-phase curve for 1/$\tau
\Delta_{0}$=26, ($\Box$) Calculated AF-phase curve incorporating
WFM. Note that $T_{WFM}$ here is chosen to be
3~K.}\label{fig:ENBCrhoWFM}
\end{figure}

As temperature further reduces below $T_{N}$, the theoretical curve
in Fig.~\ref{fig:ENBCrhoAG} overestimates the experimental curve
below 3~K ($\sim$0.3$T_{c}$). This is due to additional pairbreaking
effect of the ferromagnetic moments in the WFM phase, which shows up
as a small peak near 2.3~K (see Fig.~\ref{fig:Lambda}). The small
dip in superfluid density shows that this WFM slightly depresses,
but does not completely destroy, superconductivity, demonstrating
the coexistence of WFM and superconductivity. We model this WFM by
including a temperature-dependent magnetic impurity scattering rate
$1/\tau_{2}^{WFM} = 1/20 (g_{J}-1)J(J+1)(1-T/T_{WFM})^{\nu
^{\prime}}$ (with the same value of $g_{J}$ and $J$ as before), and
adding this to the previous effective magnetic scattering rate, i.e.
$1/\tau_{2}^{total} = 1/\tau_{2}^{eff} + 1/\tau_{2}^{WFM}$ when $T <
T_{WFM}$. The pre-factor 1/20 arises from the fact that one out of
every 20 spins contributes to the WFM \cite{Furukawa02}, giving rise
to a weak magnetization. The temperature dependence
$(1-T/T_{WFM})^{\nu ^{\prime}}$ is analogous to the molecular field
formulation. We obtain $\nu^{\prime} \approx 2$ from Jensen's
calculation \cite{Jensen02} or Choi and Canfield's data
\cite{Choi01,Canfield96}. Note that $\rho_{s}$ already starts to
flatten out at 3~K, consistent with neutron-scattering data, which
shows that this weak ferromagnetic component already shows up at 3~K
\cite{Choi01}. Hence we choose $T_{WFM}$=3~K in this WFM
calculation. Fig.~\ref{fig:ENBCrhoWFM} shows $\rho_{s}$ when one
accounts for WFM. The calculated $\rho_{s}$ does flatten out below
3~K, but does not increase below 2.3~K as the data did. There is as
yet no direct measurement of superconducting gap amplitude at this
temperature range, though our model predicts a drop in $\Delta $
there. One may need to include the effect of inelastic
spin-fluctuation scattering in this low temperature region, which
was ignored by the CN model.

In conclusion, we present in-plane penetration depth data of
single-crystal ErNi$_{2}$B$_{2}$C down to 0.1~K. The small increase
in pairbreaking at $T_{N}$ can be attributed to the interplay
between the effects of the AF molecular field and spin-fluctuation
scattering. It could also be due to the combined effects of
non-perfect nesting on one piece of the FS and non-nesting on other
pieces of the FS. The increased pairbreaking at $T_{WFM}$ is
modelled by a magnetic impurity scattering parameter, and both
magnetic orders coexist with superconductivity.

E.E.M.C wishes to thank Y. C. Chang for teaching the computational
aspects of the Chi-Nagi model, as well as C. Varma and J. Thompson
for encouraging discussions on interpretation of data. This material
is based upon work supported by the U.S. Department of Energy,
Division of Materials Sciences under Award No. DEFG02-91ER45439,
through the Frederick Seitz Materials Research Laboratory at the
University of Illinois at Urbana-Champaign. Research for this
publication was carried out in the Center for Microanalysis of
Materials, University of Illinois at Urbana-Champaign. This work in
Korea was supported by the Ministry of Science and Technology of
Korea through the Creative Research Initiative Program.

\bibliography{RNBC,CeCoIn5,Vortex}
\bigskip

\end{document}